\documentclass[latex,iicol,sn-mathphys-num]{sn-jnl}

\usepackage{graphicx}%
\usepackage{multirow}%
\usepackage{amsmath,amssymb,amsfonts}%
\usepackage{amsthm}%
\usepackage{mathrsfs}%
\usepackage[title]{appendix}%
\usepackage{xcolor}%
\usepackage{chemformula}
\usepackage{textcomp}%
\usepackage{manyfoot}%
\usepackage{booktabs}%
\usepackage{algorithm}%
\usepackage{algorithmicx}%
\usepackage{algpseudocode}%
\usepackage{listings}%
\usepackage{comment}
\usepackage{tabularx}
\usepackage{booktabs}
\usepackage{makecell}
\usepackage{geometry}

\begin{document}
\title[Article Title]{Discovery of oxide Li-conducting electrolytes in uncharted chemical space via topology-constrained crystal structure prediction}


\author[1]{\fnm{Seungwoo} \sur{Hwang}}
\author[1]{\fnm{Jiho} \sur{Lee}}

\author[1,2]{\fnm{Seungwu} \sur{Han}}

\author*[3]{\fnm{Youngho} \sur{Kang}}\email{youngho84@inu.ac.kr}

\author*[4,5]{\fnm{Sungwoo} \sur{Kang}}\email{sung.w.kang@kist.re.kr}

\affil[1]{\orgdiv{Department of Materials Science and Engineering}, \orgname{Seoul National University}, \orgaddress{\city{Seoul}, \postcode{08826}, \country{Republic of Korea}}}

\affil[2]{\orgdiv{AI Center}, \orgname{Korea Institute of Advanced Study}, \orgaddress{\city{Seoul}, \postcode{02455}, \country{Republic of Korea}}}

\affil[3]{\orgdiv{Department of Materials Science and Engineering}, \orgname{Incheon National University}, \orgaddress{\city{Incheon}, \postcode{22012}, \country{Republic of Korea}}}

\affil[4]{\orgdiv{Computational Science Research Center}, \orgname{Korea Institute of Science and Technology (KIST)}, \orgaddress{\city{Seoul}, \postcode{02792}, \country{Republic of Korea}}}

\affil[5]{\orgdiv{Division of Nanoscience and Technology, KIST School}, \orgname{University of Science and Technology (UST)}, \orgaddress{\city{Seoul}, \postcode{02792}, \country{Republic of Korea}}}

\abstract{Oxide Li-conducting solid-state electrolytes (SSEs) offer excellent chemical and thermal stability but typically exhibit lower ionic conductivity than sulfides and chlorides. This motivates the search for new oxide materials with enhanced conductivity. Crystal structure prediction is a powerful approach for identifying such candidates. However, the structural complexity of oxide SSEs, often involving unit cells with more than 100 atoms, presents significant challenges for conventional methods. In this study, we introduce TOPIC, a structure prediction algorithm that reduces configurational complexity by enforcing corner-sharing (CS) bond topology constraints. We demonstrate that TOPIC successfully reproduces the ground-state and metastable structures of known oxide SSEs, including \ch{LiTa2PO8} and \ch{Li7La3Zr2O12}, which contain up to about 200 atoms per unit cell. By combining this approach with a pretrained machine-learning interatomic potential, we systematically screen quaternary oxide compositions and identify 92 promising candidates with CS frameworks. In particular, \ch{Li4Hf2Si3O12}, which corresponds to the ground state at its composition, exhibits an ionic conductivity of 14 mS~cm$^{-1}$, a hull energy of 21 meV~atom$^{-1}$, and a band gap of 6.5 eV. Through our investigation, we identify the Li ratio as one of the key factors determining the stability of CS structures. Overall, our approach provides a practical and scalable pathway for discovering high-performance oxide solid electrolytes in previously unexplored chemical spaces.}



\maketitle

\clearpage 

\section{Introduction}\label{sec1}

Despite extensive research into lithium-ion battery technologies, persistent safety issues and limited capacity continue to impede their widespread adoption in high-energy applications, such as electric vehicles and large-scale energy storage.\cite{EES_review_2018, JPC_viewpoinit_2015, NatMater_review_2017,fire_report_1994, Thermalrunaway_2012} Solid-state electrolytes (SSEs) offer an attractive solution by supporting stable operation at higher voltages.\cite{Lau_AEM_review_2018, Chen_Chem_review_2020} Among SSE candidates, oxide-based electrolytes are particularly promising due to their superior chemical and thermal stability.\cite{Kim_AEM_review_2021} Several oxide SSEs, such as garnet-type \ch{Li7La3Zr2O12} (LLZO),\cite{Murugan_LLZO_2007} perovskite-type Li$_{3x}$La$_{\left(2/3 - x\right)}$TiO$_{3}$ (LLTO),\cite{Stramare_LLTO_2003} NASICON-type Li$_{1.3}$Al$_{0.3}$Ti$_{1.7}$(PO$_{4}$)$_{3}$ (LATP),\cite{Aono_NASICON_1990} and  \ch{LiTa2PO8}\cite{Kim_LiTa2PO8_2018} have achieved ionic conductivities above 1 mS~cm$^{-1}$. However, their ionic conductivities are still generally lower than those of sulfide materials,\cite{EES_review_2018} highlighting the need to discover novel oxide SSEs.

Computational screening can accelerate discovery of new SSE materials in much higher speed compared to experiments. In fact, there have been efforts to discover SSEs by searching the existing database, such as Inorganic Crystal Structure Database (ICSD),\cite{Bergerhoff_ICSD_1983} have successfully identified promising candidates.\cite{Mo_AEM_screening_2019, Marzari_screening_2020, He_SPSE_screening_2020} To further expand the search space, several studies have employed structures available in existing databases as templates, systematically substituting their constituent elements to generate new candidate materials. \cite{Sendek_HT_screening_2016,Fujimura_LISICON_screening_2013, Zhu_LPS_LMPS_screening_2017, Kim_Mok_Kim_Back_2023} However, these data-driven schemes inherently lack the capability to uncover new structural prototypes, posing a critical challenge for exploring uncharted chemistries. \cite{Oganov_Pickard_Zhu_Needs_2019}

Discovery of materials with novel structural prototypes requires crystal structure prediction, which is a computational approach used to identify ground-state or metastable crystal structures for a given chemical composition.\cite{Oganov_USPEX_2006, Call_PSO_2007, Hwang_JACS_2023} These methods have successfully led to the discovery of diverse material applications, such as the superhard materials,\cite{Oganov_gamma_boron_2009} high-temperature superconductors,\cite{Duan_high_Tc_CSP_2014} and two-dimensional electrides.\cite{Ming_electride_CSP_2016} Crystal structure prediction is also used to discover novel SSE materials.\cite{Wang_LiAlSO_CALYPSO_2017, Wang_Li3SmCl6_CALPYSO_2024, Zhang_CSP_SSE_2025, Rosseinsky_SSE_CSP_2025} However, the discovery of oxide SSEs via crystal structure prediction remains significantly more challenging than for other types, such as sulfides or chlorides. This is because oxide SSEs often possess highly complex structures, typically containing over 100 atoms per unit cell and involving quaternary or higher-order compositions.\cite{EES_review_2018} Conventional structure prediction methods, including genetic algorithms, are generally restricted to simpler systems because of the high computational cost for evaluating energies of intermediate and candidate structures using density functional theory (DFT) calculations. To address this limitation, several studies have employed machine-learning interatomic potentials (MLIPs)\cite{Behler_Parrinello_2007, Shapeev_MTP_2016, Bartók_Payne_Kondor_Csányi_GAP_2010}—surrogate models trained on DFT data to predict energies and forces—in crystal structure prediction, as they offer significantly higher speed and comparable accuracy to DFT calculations.\cite{MTP_USPEX_2019, MQA_CSP_2020, GAP_CALYPSO_2018, GAP_RSS_2018} However, even when accelerated by MLIPs, reported crystal structure prediction studies have thus far been limited to ternary systems with fewer than 50 atoms per unit cell, or to multinary systems with high-symmetry structures,\cite{Kang_SPINNER_2022, Han_NCS_efficient_CSP_2025} which remains insufficient for the discovery of oxide SSE materials. Therefore, more efficient prediction methods are needed to address the complexity of oxide SSE structures.

Using structural constraints in crystal structure prediction algorithms can further accelerate the process, especially when prior knowledge of structural characteristics (such as the shapes of known polyanions) is available.\cite{Sato_SSE_CRYSPY_2024, Zhang_CSP_SSE_2025} In the case of oxide SSEs, it is well known that most existing materials exhibit oxygen-corner-sharing (CS) frameworks, whereas sulfides and halides typically include isolated units such as \ch{PS4}. Based on this knowledge, experimental efforts targeting compositions likely to yield CS frameworks led to the discovery of \ch{LiTa2PO8}, which was indeed confirmed to exhibit a CS framework.\cite{Kim_LiTa2PO8_2018} Furthermore, computational screening of the Materials Project database\cite{Materials_project_2013} focusing on CS-framework structures successfully identified oxide materials with high ionic conductivity, including \ch{LiGa(SeO3)2}, which is also validated through experiments.\cite{Ceder_CS_2022} In that study, the superior ionic conductivity of CS frameworks was rationalized through theoretical analysis, providing insight into their structure–property relationships.

Here, leveraging the knowledge that oxide materials tend to exhibit CS frameworks, we develop a crystal structure prediction algorithm named TOPIC (\textbf{TO}pology-constrained \textbf{P}rediction of \textbf{I}norganic \textbf{C}rystals). TOPIC generates candidate structures under CS bond topology constraints, significantly reducing the configurational search space compared to conventional methods. We first validate the TOPIC algorithm by successfully reproducing the crystal structures of known oxide systems containing up to 192 atoms per unit cell, including LLZO. Next, using a pretrained MLIP (SevenNet-0),\cite{Park_SevenNet_2024} we systematically screen for novel quaternary oxide SSEs exhibiting CS frameworks generated by TOPIC. Our initial screening focuses on chemical systems composed of commonly used elements, such as Ti, Zr, and P, where we find that the lithium content serves as a key descriptor for the stability of CS frameworks. Based on this descriptor, we enumerate possible quaternary compositions likely to exhibit CS frameworks composed of octahedral and tetrahedral units and apply TOPIC to explore such chemical space. Overall, we identify 92 new candidate materials with potentially high ionic conductivity. Among them, 15 oxide materials are further validated using DFT calculations to assess their thermodynamic stability, electronic band gaps, and ionic conductivity. Finally, we report several novel framework types with high predicted ionic conductivity and derive design principles by analyzing the predicted SSE structures and the Li-ion conduction pathways.

\begin{figure*}[h]
 \centering
 \includegraphics[width=12cm]{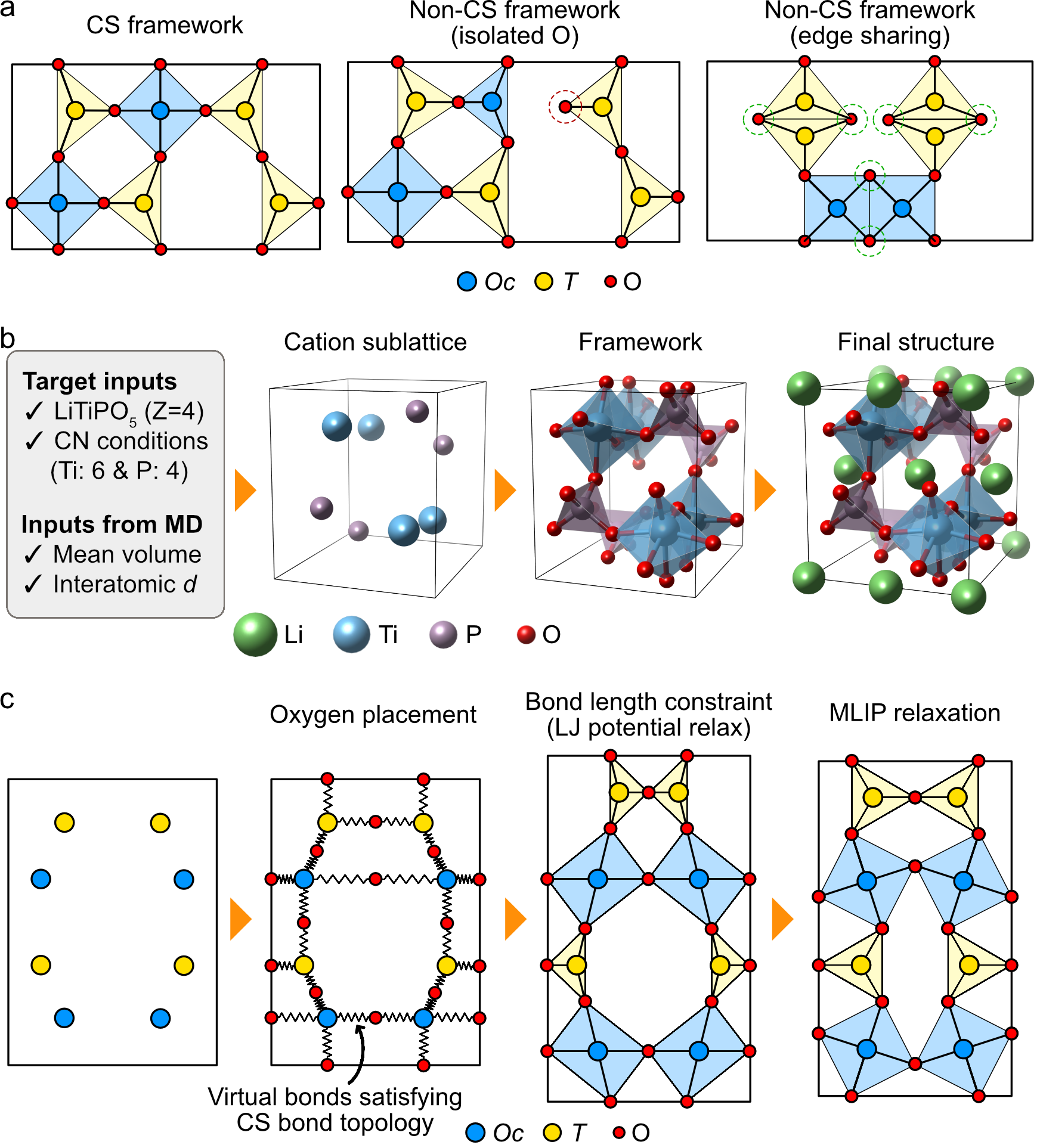}
 \caption{TOPIC algorithm. (a) Schematic illustrations of fully connected corner-sharing framework (left), framework with isolated vertex (middle), and framework with edge-sharing polyhedra (right). Dashed red circle in the middle panel indicates the isolated vertex in framework, and dashed green circles in right panel indicate nodes on edge-shared parts. \textit{Oc} and \textit{T} represent the cations form octahedral and tetrahedral polyhedra, respectively. (b) Overview of TOPIC algorithm. \ch{LiTiPO5} case is shown as an example. (c) Detail schematic process of cation-oxygen framework generation. Spring lines in the second panel indicate the virtual bonds between cation--oxygen pairs to which the Lennard--Jones potential is applied.}
 
 \label{Fig1}
\end{figure*}

\section{Results}\label{sec2}
\subsection{Definition of corner-sharing bond topology}
We define a bond topology as a connectivity of a three-dimensional network formed by polyhedra centered on cations (except Li) with surrounding oxygen atoms. A fully connected CS framework is one in which every oxygen bridges exactly two polyhedra, thereby establishing continuous connectivity (left panel of Fig.~\ref{Fig1}a). This definition explicitly excludes configurations where an oxygen anion binds to only one polyhedron, forming an isolated vertex (middle panel of Fig.~\ref{Fig1}a), or configurations where two polyhedra share an edge or face (right panel of Fig.~\ref{Fig1}a). Our definition thus slightly differs from the broader criterion proposed by Jun et al.,\cite{Ceder_CS_2022} which allows isolated vertices. Hereafter, we refer to frameworks with fully connected vertex-sharing topologies as CS frameworks, and we classify all other cases as non-CS frameworks.

\subsection{TOPIC algorithm}

We develop TOPIC, a random structure generation algorithm constrained by CS bond topology. Structural optimizations and energy evaluations are carried out using MLIPs during the structure search, and the final candidates are recalculated with DFT to obtain accurate energies. To construct the training set for MLIPs, we perform melt--quench--annealing molecular dynamics (MD) simulations at a target SSE composition to generate disordered structures using the SPINNER code,\cite{Kang_SPINNER_2022} following a procedure similar to our previous works.\cite{MQA_CSP_2020, Hwang_JACS_2023} From this training set, we train three types of MLIPs for each composition: (1) a 3~\AA\ cutoff model excluding Li atoms, (2) a 6~\AA\ cutoff model excluding Li atoms, and (3) a 6~\AA\ cutoff model including all atoms. The first two models are trained by removing Li atoms from the configurations while retaining the original energy and force labels for the others, which include the contribution from Li. In other words, the energetic contribution of Li is implicitly averaged out during training. These three MLIPs are then used in the structural search. Interatomic distances (such as cation–-oxygen bond lengths) are extracted from the MD trajectories for generating training set and used as geometric constraints during structure generation (see Methods for details).

TOPIC generates random CS frameworks without Li and subsequently determines the optimized Li positions. Specifically, the algorithm sequentially performs following steps: (1) generating cation sublattices, (2) generating cation--O framework by placing oxygen atoms under bond topology constraints followed by relaxations, and (3)  placing Li atoms (see Fig.~\ref{Fig1}b). At the first step, the cation sublattices are generated using random structure generation with space group constraints using RandSpg code,\cite{Avery_randSpg_2017} as similar methods are commonly adopted in other structure prediction algorithms.\cite{Oganov_USPEX_2006, Pickard_AIRSS_origin_2006, Pickard_AIRSS_2011} At the final step, Li atoms are placed via Monte Carlo (MC) simulations coupled with MLIP-based optimizations. In this approach, virtual bonds are first formed between selected cation pairs on the given cation sublattice, and oxygen atoms are placed at the midpoints of these bonds (see Fig.~\ref{Fig1}c). The resulting network is required to satisfy the CS bond topology and the predefined coordination number for each cation species, as determined by Pauling’s rules~\cite{Pauling_1929} and Shannon’s ionic radii~\cite{Shannon_1976} (Table~S1). (We also confirm that O atoms are located near the midpoint between cation pairs in experimentally known materials; see Fig.~S1.) The resulting structures are optimized by applying a Lennard-Jones (LJ) potential to the virtual bonds to enforce reasonable cation–-oxygen bond lengths. In addition, a harmonic repulsive potential is used to prevent unphysically short distances and unintended increases in coordination numbers during optimizations. Finally, the structures are further relaxed using MLIP optimizations. After each relaxation step (both LJ and MLIP), we verify whether the structure still satisfies the CS bond topology and the predefined coordination numbers. Structures that violate these criteria are discarded. Further details of each algorithmic step are provided in the Methods section.

\begin{figure*}[!t]
 \centering
 \includegraphics[height=17cm]{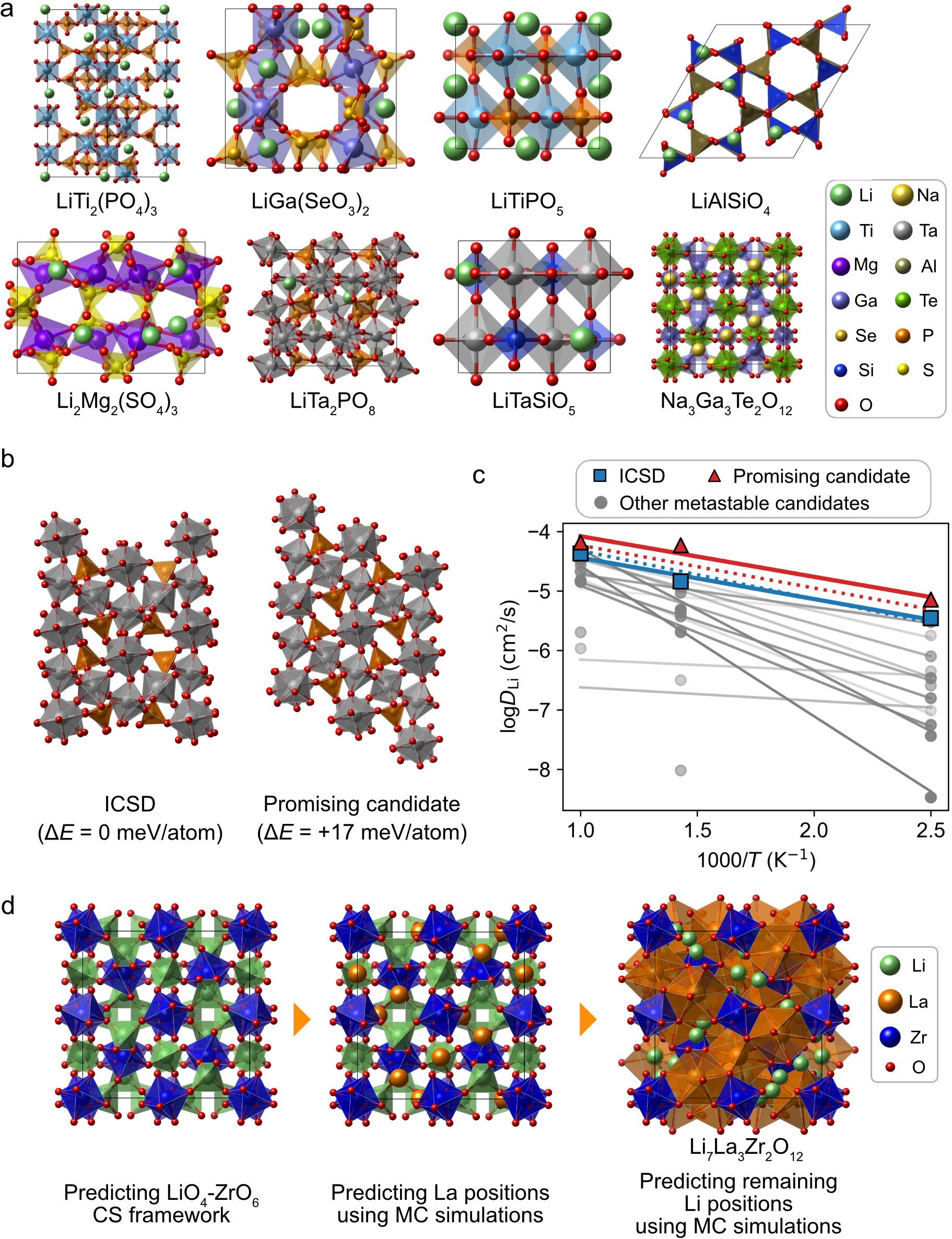}
 \caption{Validation of the TOPIC algorithm. (a) The lowest-energy structures for each validation test system generated by TOPIC. (b) The lowest-energy and a new metastable structure in the \ch{LiTa2PO8} system obtained by TOPIC. (c) Diffusion coefficients at 400, 700, and 1000 K for all polymorphs of the \ch{LiTa2PO8} system obtained by TOPIC. Blue lines represent the known structure, the red line corresponds to the structure of the promising candidate, and gray lines indicate other polymorphs. Solid lines denote results calculated by MD simulations using SevenNet-0, while dashed lines denote results obtained by DFT simulations. (d) Revised structure generation process for cubic \ch{Li7La3Zr2O12}. Left: the framework consisting of \ch{ZrO6} octahedra and \ch{LiO4} tetrahedra generated by TOPIC. Middle: the framework after La atoms are introduced via Voronoi tessellation and Monte Carlo simulation. Right: the final \ch{Li7La3Zr2O12} structure after the Li insertion.}
 \label{Fig2}
\end{figure*}

In the CS-framework-generation step, structural optimizations are performed using MLIPs trained on datasets that exclude the effects of Li atoms, as described above. With a cutoff radius of 6~\AA\, these MLIPs yield average validation errors of 0.014 eV~atom$^{-1}$ for energy and 0.87 eV~\AA$^{-1}$ for force. More accurate calculations are subsequently conducted with MLIPs trained including Li, during the Li-insertion stage. This two-step optimization procedure (from Li-free to Li-occupied systems) effectively identifies structures that satisfy CS-topology constraints while reducing computational cost, as demonstrated in the following subsection. The effectiveness of the initial framework-generation step arises from the intrinsically weak Li–O interactions in multicomponent Li oxides, which result in only minor modifications of the Li-free structural framework upon Li insertion. Weak Li–O interactions have been reported in NASICON-type Li compounds, which exhibit low-frequency phonon modes and notably weak Li–O bonding, as confirmed by crystal orbital Hamilton population (COHP) analyses.\cite{Böger_NASICON_COHP_2024, Deringer_COHP_2011} Our COHP analysis similarly confirms weak Li–O interactions in compounds such as \ch{LiTi2(PO4)3}, \ch{LiGa(SeO3)2}, and \ch{LiTa2PO8}, as well as in the melt-quenched amorphous phases of the same compositions (Fig.~S2). 

\begin{table}
\small
  \caption{\ Information of validation test systems. Space group, number of atoms in unit cell ($N_{\mathrm{atom}}$), and the number of discovered reference structures in 300,000 trials ($N_{\mathrm{found}}$) are provided.}
  \label{Table1}
  \begin{tabular*}{0.48\textwidth}{@{\extracolsep{\fill}}llll}
    \hline
    Formula & Space group & $N_{\mathrm{atom}}$ & $N_{\mathrm{found}}$\\
    \hline
    \ch{LiTi2(PO4)3} & R-3c (167) & 108 & 7620\\
    \ch{LiGa(SeO3)2} & I-42d (122) & 80 & 24\\
    \ch{Li2Mg2(SO4)3} & Pbcn (60) & 76 & 2\\
    \ch{LiTa2PO8} & C2/c (15) & 96 & 2\\
    \ch{Na3Ga3Te2O12} & Ia-3d (230) & 160 & 10575\\
    $\alpha$-\ch{LiTiPO5} & P-1 (2) & 32 & 1\\
    $\beta$-\ch{LiTiPO5} & Pnma (62) & 32 & 155\\
    $\alpha$-\ch{LiAlSiO4} & R3 (148) & 126 & 148\\
    $\beta$-\ch{LiAlSiO4} & P$6_{4}$22 (181) & 84 & 12\\
    $\gamma$-\ch{LiAlSiO4} & Pc (7) & 28 & 1\\
    \ch{LiTaSiO5} & P$2_{1}$/c (14) & 32 & 76\\
    \ch{Li7La3Zr2O12} & Ia-3d (230) & 192 & 10512\\
    \hline
  \end{tabular*}
\end{table}

\subsection{Validation of TOPIC}

To assess the reliability of TOPIC, we check whether it reproduces the structures of known materials. The test set includes three experimentally reported compositons (\ch{LiTi2(PO4)3}, \ch{LiAlSiO4}, and \ch{LiTa2PO8}); four computationally proposed candidates (\ch{LiTaSiO5}, \ch{LiTiPO5}, \ch{Li2Mg2(SO4)3}, and \ch{LiGa(SeO3)2});\cite{Mo_LiAlSiO4_LiTaSiO5_2017, Ceder_CS_2022} and one garnet-type Na compound with corner-sharing bond topology (\ch{Na3Ga3Te2O12}). For the two compositions, \ch{LiTiPO5} and \ch{LiAlSiO4}, which exhibit multiple experimentally observed phases, we examine whether TOPIC accurately reproduces all of these phases. Specifically, \ch{LiTiPO5} exhibits $\alpha$ (P-1) and $\beta$ (Pnma) phases,\cite{Hupfer_LiTiPO5_phase_transition_2017} while \ch{LiAlSiO4} has three known phases: $\alpha$ (R3), $\beta$ (P$6_{4}$22), and $\gamma$ (P1c1).\cite{Norby_LiAlSiO4_phase_transition1990} Therefore, total eleven structures are accounted for in the test. We note that \ch{LiTa2PO8} exhibits disordered Li sites, whereas the other compounds have fixed Li sites. A summary of test systems and search outcomes is provided in Table~\ref{Table1}.

\begin{figure*}[h]
 \centering
 \includegraphics[height=14cm]{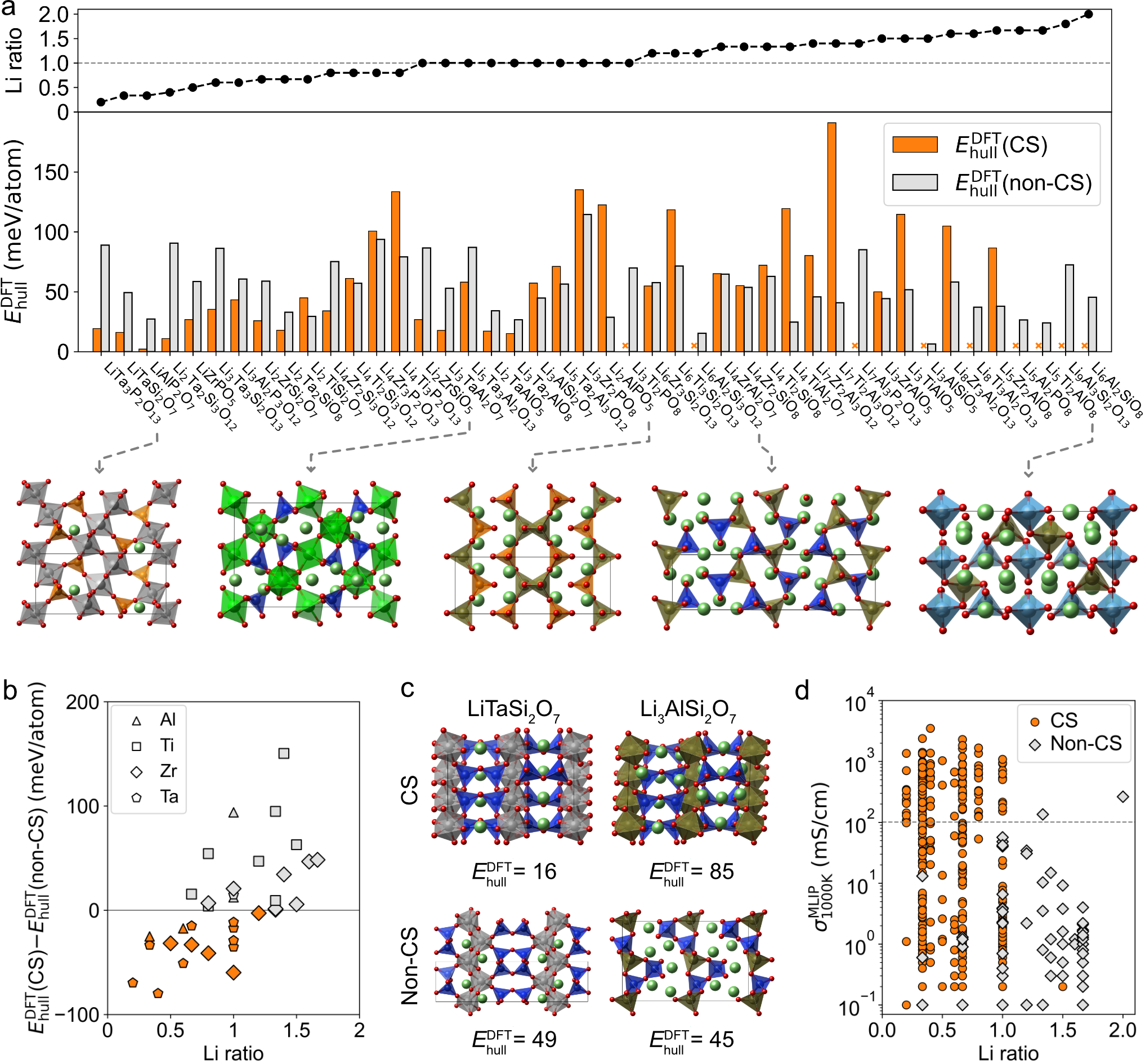}
 \caption{Structural stability of oxide SSEs with CS bond topology in representative element sets. (a) Upper part: Li ratio of the searched materials. Lower part: $E_{\mathrm{hull}}^{\mathrm{DFT}}$ of CS and non-CS frameworks in each composition. Compositions where no stable CS structure is found are represented by x markers. The lowest-energy atomistic structures of \ch{LiTa3P2O13}, \ch{Li2ZrSiO5}, \ch{Li2AlPO5}, \ch{Li6Al2Si3O12}, and \ch{Li5Ti2AlO8} are presented below the bar plot as example structures at each Li ratio. (b) $E_{\mathrm{hull}}^{\mathrm{DFT}}$(CS) $-$ $E_{\mathrm{hull}}^{\mathrm{DFT}}$(non-CS) as a function of Li ratio. (c) Structures of CS and non-CS framework in \ch{LiTaSi2O7} and \ch{Li3AlSi2O7} and their energy above hull values (unit: meV atom$^{-1}$). (d) Ionic conductivity at 1000 K calculated by SevenNet-0 as a function of Li ratio for both CS and non-CS framework structures.}
 \label{Fig3}
\end{figure*}

For each system, we perform 300,000 trials and identify the lowest-MLIP-energy structure, followed by DFT evaluations of candidates within a 30 meV~atom$^{-1}$ window of this lowest-energy structure. Note that the number 300,000 refers to the structures initially generated to satisfy the CS bond topology, prior to any optimization with the LJ potential or MLIPs. After optimization, only a small fraction of these 300,000 structures remain (approximately 7\%), preserving the CS framework (see Methods for details).


Fig.~\ref{Fig2}a presents the lowest-energy structures (within DFT calculations) obtained for each system, and comparisons with ICSD reference structures (including metastable polymorphs) are shown in Fig.~S3. Among ten target structures with ordered Li sites, TOPIC successfully predicts six with the correct Li positions: \ch{LiTi2(PO4)3}, \ch{LiGa(SeO3)2}, $\beta$-\ch{LiTiPO5}, $\alpha$-\ch{LiAlSiO4}, $\beta$-\ch{LiAlSiO4}, and \ch{Na3Ga3Te2O12}. For \ch{Li2Mg2(SO4)3}, the predicted framework reproduces the known structure with only minor deviations in Li sites, yet all predicted Li positions remain within established conduction pathways (Fig.~S3g). For \ch{LiTaSiO5}, $\alpha$-\ch{LiTiPO5}, and $\gamma$-\ch{LiAlSiO4}, TOPIC predicts slightly distorted structures, while the polyhedral connectivities remain consistent with the reference structures. In the case of \ch{LiTa2PO8}, the reference structure contains partially occupied Li sites; here, we confirm that the predicted framework is identical to the reference and that the predicted Li positions are on the conduction channels (Fig.~S3h).

TOPIC also predicts low-energy structures different from those previously reported. For example, in \ch{LiTaSiO5}, it identifies a structure that is 6 meV~atom$^{-1}$ lower in energy than the reported phase, distinguished by a different stacking sequence (Fig.~S4a). In addition, we discover a previously unreported metastable polymorph of \ch{LiTa2PO8}, which lies 17 meV~atom$^{-1}$ higher in energy than the reported phase. This polymorph differs from the known phase in its stacking sequence but retains similar polyhedral motifs (Fig.~\ref{Fig2}b and Fig.~S4b). Note that this new polymorph of \ch{LiTa2PO8} exhibits nearly twice the conductivity of the previously known phase (Fig.~\ref{Fig2}c, dashed line).

Finally, we discuss LLZO, which is the only oxide SSE known not to exhibit a CS framework. However, if one considers a framework built from \ch{ZrO6} octahedra and \ch{LiO4} tetrahedra instead of focusing on Zr and La, the cubic phase of LLZO can be interpreted as a corner-sharing framework (i.e., \ch{Li4La3(ZrO6/2)2(LiO4/2)3}).\cite{Kim_LiTa2PO8_2018} We slightly modify the TOPIC algorithm (see Methods for details) to examine whether it can reproduce the structure of LLZO (see Fig.~\ref{Fig2}d). Specifically, we first predict the \ch{ZrO6}-\ch{LiO4} framework, then place La atoms using MC simulations. Finally, we remove the pre-existing Li atoms and re-predict all Li positions with MC simulations. From this modified process, we successfully identify the LLZO structure, thereby confirming the effectiveness of the TOPIC algorithm. In the screening process outlined in the following subsections, La is excluded from the search space because the method requires modification. Future research, however, could systematically extend the TOPIC algorithm to this material family, as demonstrated above.


\subsection{Structure exploration in representative Li-SSE element sets}


We first explore candidate materials composed of cation polyhedra with octahedral and tetrahedral coordination, with general stoichiometry Li$_x$($\mathit{Oc}$O$_{6/2}$)$_m$($\mathit{T}$O$_{4/2}$)$_n$. Here, the lithium content ($x$) is determined by charge neutrality, and $\mathit{Oc}$ and $\mathit{T}$ denote cations occupying octahedral and tetrahedral sites, respectively. Based on their frequent occurrence in oxide SSE frameworks,\cite{Kim_AEM_review_2021, Umair_review_2024} we select Ta, Ti, Zr, and Al for $\mathit{Oc}$, and P, Si, and Al for $\mathit{T}$. Considering $\mathit{Oc}$:$\mathit{T}$ ratios of 2:1, 3:2, 1:1, 2:3, and 1:2—the most common cation ratios observed in Li-containing quaternary oxides (see Fig.~S5)—we generate 50 candidate compositions in total. From this set, we direct our attention to new compositions that have not yet been explored experimentally and are absent from ICSD.\cite{Bergerhoff_ICSD_1983} After removing compositions that cannot satisfy charge neutrality for $x > 0$, 44 unique compositions remain as our target space. For each composition, TOPIC performs 300{,}000 trials per formula unit ($Z$), with $Z$ = 4, 6, and 8 for $\mathit{Oc}$:$\mathit{T}$ ratios of 2:1, 1:1 and 1:2, and $Z$ = 2, 4, and 6 for $\mathit{Oc}$:$\mathit{T}$ ratios of 2:3 and 3:2, resulting in unit cells containing up to about 100 atoms. For comparison, we apply SPINNER, an evolutionary algorithm accelerated by MLIPs,~\cite{MQA_CSP_2020, Kang_SPINNER_2022, Hwang_JACS_2023} without topology constraints, to generate structures beyond the CS topology. SPINNER generates 60{,}000 candidate structures at $Z = 4$ for $\mathit{Oc}$:$\mathit{T}$ ratios of 2:1, 1:1, and 1:2, and $Z = 2$ for ratios of 2:3 and 3:2, resulting in unit cells with approximately 50 atoms to maintain computational feasibility. Following structural generation in both methods, DFT energies are computed for candidates lying within 50 meV~atom$^{-1}$ of the lowest-MLIP-energy structure.

By comparing the lowest-energy structures obtained from each method, we calculate the energies above the convex hull ($E_{\mathrm{hull}}^{\mathrm{DFT}}$) for both CS and non-CS frameworks, hereafter denoted as $E_{\mathrm{hull}}^{\mathrm{DFT}}$(CS) and $E_{\mathrm{hull}}^{\mathrm{DFT}}$(non-CS), respectively. Fig.~\ref{Fig3}a shows $E_{\mathrm{hull}}^{\mathrm{DFT}}$(CS) and $E_{\mathrm{hull}}^{\mathrm{DFT}}$(non-CS) across all compositions studied in the present work. Note that experimentally synthesized SSEs often exhibit positive $E_{\mathrm{hull}}^{\mathrm{DFT}}$ values (e.g., \ch{LiTa2PO8} (26 meV~atom$^{-1}$) and \ch{LiNbOCl4} (33 meV~atom$^{-1}$)).\cite{Hussain_LiTa2PO8_hullE_2019, Tanaka_oxyhalide_SSE_2023} In comparison, the SSEs identified in this work are also exhibit positive but comparable or smaller $E_{\mathrm{hull}}^{\mathrm{DFT}}$ values, suggesting their potential synthesizability and stability. We find a correlation between the Li ratio and the relative stability of CS versus non-CS frameworks, where the Li ratio is defined as the number of Li atoms divided by the number of non-Li metal atoms. Systems that prefer CS frameworks over non-CS frameworks generally exhibit small Li ratios ($\leq 1.0$). Fig.~\ref{Fig3}b illustrates this trend more clearly, showing a positive correlation between the Li ratio and $E_{\mathrm{hull}}^{\mathrm{DFT}}$(CS) $-$ $E_{\mathrm{hull}}^{\mathrm{DFT}}$(non-CS). Even for several compositions with Li ratios greater than 1.0, no stable CS-topology structures are found (× markers in Fig.~\ref{Fig3}a).

One might question our conclusion regarding the higher preference for CS frameworks in systems with low Li contents, since TOPIC explores CS-topology configurations more exhaustively than non-CS configurations in SPINNER, which is limited to searching unit cells with fewer atoms than those accessible in TOPIC. However, the same conclusion is reached when comparing CS and non-CS structures generated solely by SPINNER, as shown in Fig.~S6 and Fig.~S7, although some of the lowest-energy CS structures differ from those found in TOPIC (usually, TOPIC identifies more stable CS configurations than SPINNER). This supports that the preference for CS bond topology indeed exists in systems with low Li contents. The preference for CS frameworks over non-CS ones at low Li ratios can be rationalized by the spatial constraints inherent to fully connected CS frameworks with compact atomic packing, which limit the accommodation of Li ions. On the other hand, structures with high Li content in well-known sulfide and chloride SSEs typically include isolated polyhedra (e.g., PS$_4$), providing greater internal space for Li accommodation.\cite{Deiseroth_LPSC_2008, Kamaya_LGPS_2011, Asano_LYC_2018} This is exemplified by comparing \ch{LiTaSi2O7} (low Li ratio) and \ch{Li3AlSi2O7} (high Li ratio). As shown in Fig.~\ref{Fig3}c, both structures share the same CS framework. For \ch{Li3AlSi2O7}, however, the CS polymorph is higher in hull energy than its non-CS counterpart, while for \ch{LiTaSi2O7}, the CS polymorph remains more stable. Additionally, in Fig.~\ref{Fig3}b, frameworks incorporating larger Zr ions (86 pm) consistently exhibit lower $E_{\mathrm{hull}}^{\mathrm{DFT}}$(CS) $-$ $E_{\mathrm{hull}}^{\mathrm{DFT}}$(non-CS) values than analogous frameworks containing smaller Ti ions (75 pm). This trend can be attributed to the greater internal free space in the former, arising from the larger ionic radius of $\mathrm{Zr}^{4+}$ compared to $\mathrm{Ti}^{4+}$ (further illustrated in Fig.~S8).


We examine the ionic conductivities of all lowest-energy and metastable structures within 50 meV~atom$^{-1}$ above the hull at 1000 K. To this end, we perform MLIP-MD simulations using SevenNet-0, which has been shown to predict Li-ion conductivities with reasonable accuracy (see Fig.~S9 and Table~S2 for details).\cite{Kim_FT_SevenNet_2025} Fig.~\ref{Fig3}d shows the conductivity as a function of the Li ratio. Among 339 CS-framework structures examined, 133 (39\%) exhibit conductivities above 101 mS~cm$^{-1}$, which corresponds to practically relevant values of 0.1 mS~cm$^{-1}$ at room temperature, assuming an activation energy of 0.3 eV.\cite{Ceder_CS_2022, EES_review_2018} In contrast, only 2 out of 70 non-CS structures (3\%) exceed this threshold. As expected, frameworks with isolated vertices exhibit stronger Li–O interactions and hence lower ionic conductivity—consistent with the generally poorer transport of oxide frameworks with isolated vertices (e.g., oxide LISICON-type) relative to CS framework (e.g., NASICON-type).\cite{Tao_AFM_ionic_conductivity_compare_2022} Overall, low Li concentration favors the formation of CS-framework structures, which exhibit significantly higher Li-ion conductivity compared to non-CS structures. Therefore, we suggest that the Li ratio can serve as a key descriptor for screening oxide SSE materials.

\subsection{Comprehensive screening of quaternary compositions}

\begin{figure}[h]
\centering
  \includegraphics[width=7cm]{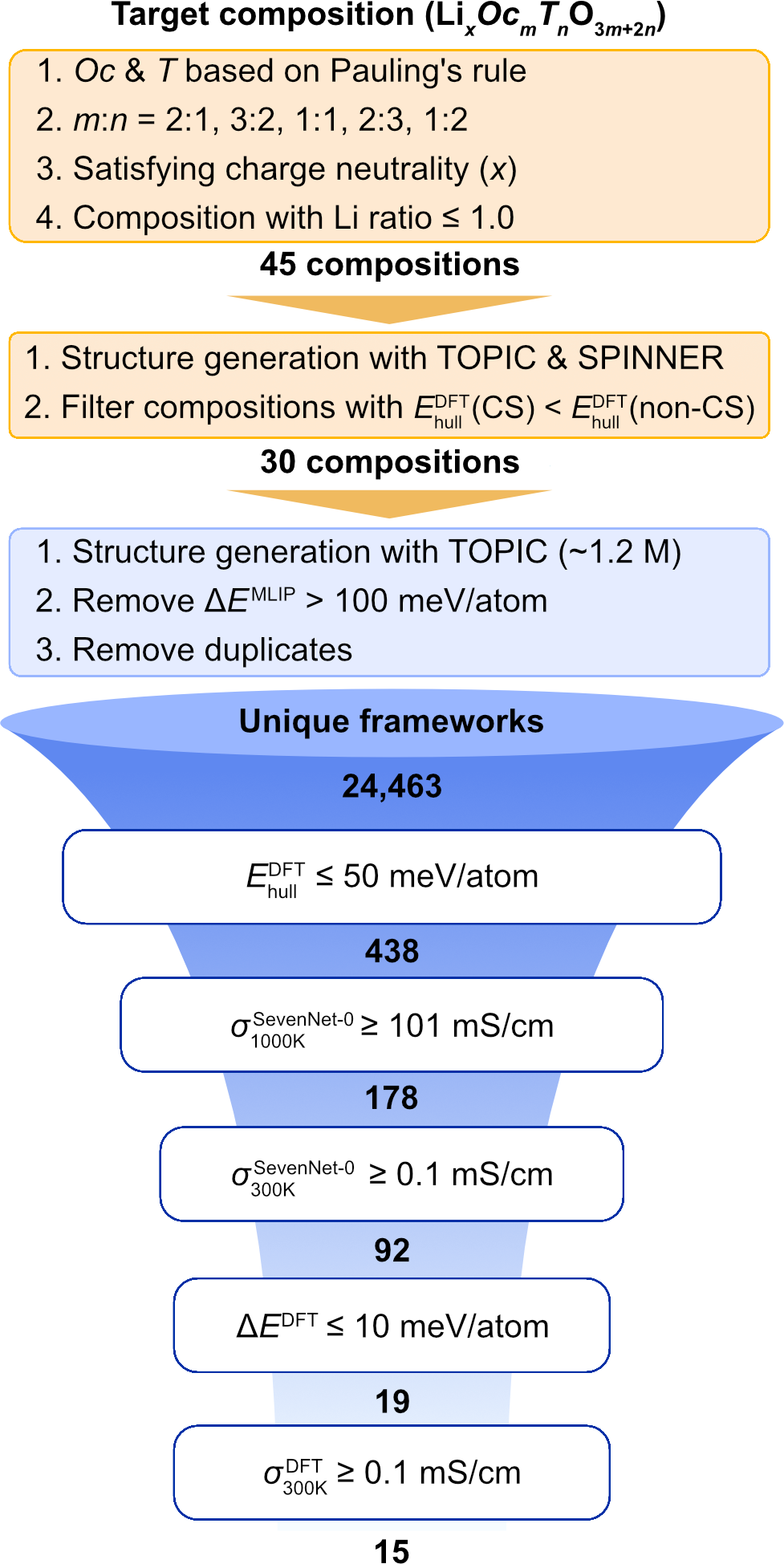}
  \caption{Screening process. Orange boxes indicate target composition selection. Unique frameworks from these compositions are then evaluated as candidate solid electrolytes under several screening conditions (blue boxes).}
  \label{Fig4}
\end{figure}

\begin{figure*}[h]
\centering
  \includegraphics[width=14.5cm]{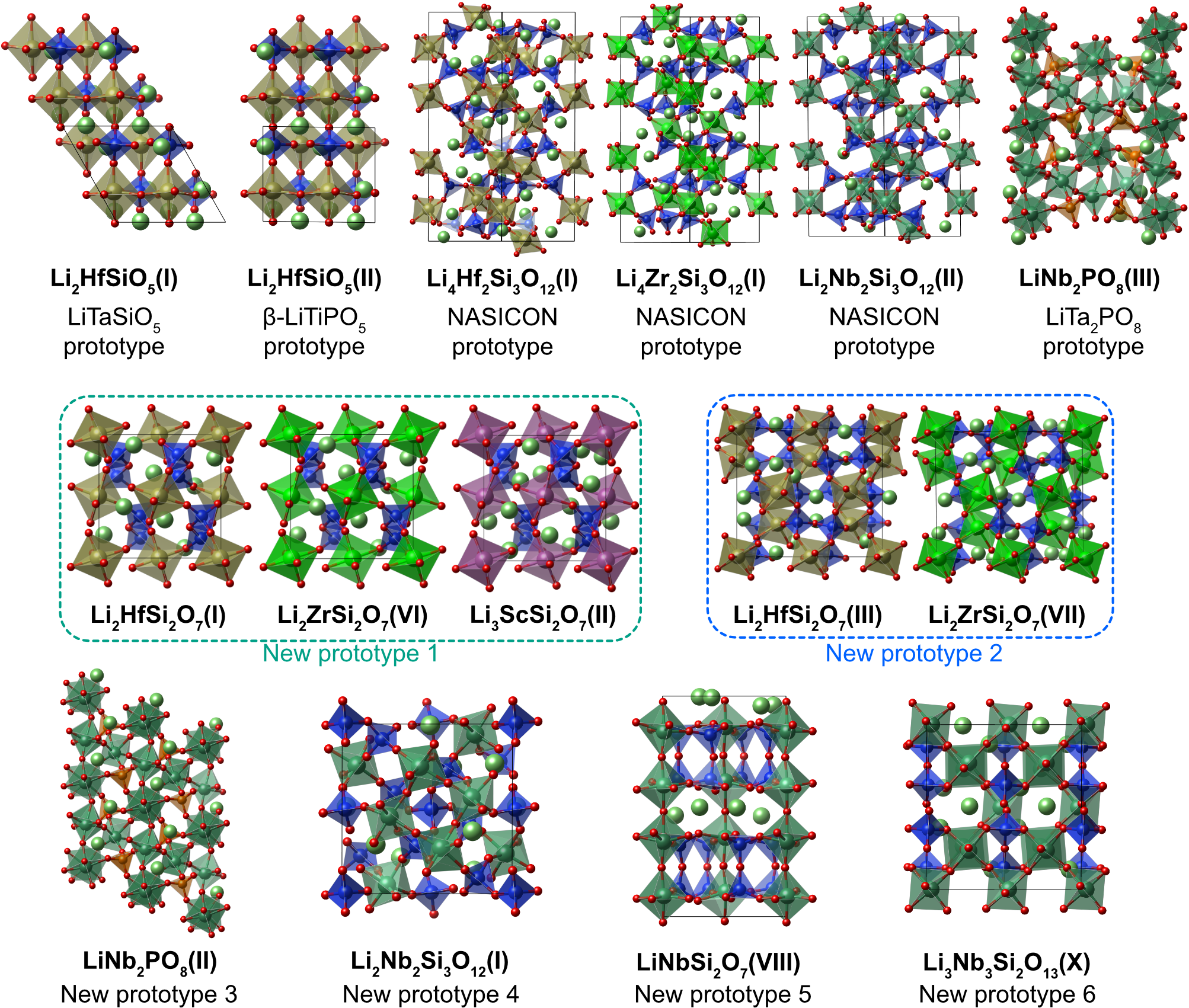}
  \caption{Final candidates. Structures are selected based on the criteria of high ionic conductivity ($\geq$0.1 mS cm$^{-1}$) and a low energy difference ($\leq$10 meV atom$^{-1}$) from the lowest-energy structure at each composition. The corresponding properties are listed in Table~2.}
  \label{Fig5}
\end{figure*}

We conduct a thorough screening of oxide SSE candidates in the quaternary compositional space using the Li-ratio descriptor. We first enumerate compositions of the form Li$_x$($\mathit{Oc}$O$_{6/2}$)$_m$($\mathit{T}$O$_{4/2}$)$_n$, considering $\mathit{Oc}$:$\mathit{T}$ ratios of 2:1, 3:2, 1:1, 2:3, and 1:2, under the constraint that the Li ratio is less than or equal to 1.0. The $Oc$ and $T$ elements discussed in the previous section are included in this screening. Furthermore, Mg, Ga, Sc, Hf, and Nb, which fulfill Pauling’s rule yet have been relatively underexplored, are included as $Oc$ elements. This procedure yields a total of 45 quaternary compositions. For each composition, we apply both TOPIC and SPINNER and retain the lower-DFT-energy structure as described above. We observe a similar correlation between the Li ratio and $E_{\mathrm{hull}}^{\mathrm{DFT}}$(CS) $-$ $E_{\mathrm{hull}}^{\mathrm{DFT}}$(non-CS) (Fig.~S10 and Table~S3) as in the previous screening results (Fig.~\ref{Fig3}b), indicating that the Li ratio descriptor is robust for these systems as well. 


To efficiently identify candidates with high ionic conductivity, we employ a stepwise screening procedure, as illustrated in Fig.~\ref{Fig4}. First, we identify 45 quaternary Li-oxide compositions with Li ratios below 1.0, as described above. Next, we perform structure predictions with SPINNER ($Z=2$ or 4, yielding unit cells with approximately 50 atoms, 60,000 structure generations) and TOPIC ($Z=4, 6$, 300,000 trials for each). We retain only those with $E_{\mathrm{hull}}^{\mathrm{DFT}}$(CS) $<$ $E_{\mathrm{hull}}^{\mathrm{DFT}}$(non-CS), reducing the set from 45 to 30. Subsequently, CS-topology crystal structures are generated with TOPIC from an expanded pool of 1.2 million trials with $Z=2,4,6$, and 8 (300,000 trials for each), enabling more reliable identification of energetically favorable configurations. After discarding structures lying more than 100 meV~atom$^{-1}$ above the lowest MLIP-energy for each composition and removing duplicates, we obtain 24,463 unique CS-framework structures. Finally, selecting those with $E_{\mathrm{hull}}^{\mathrm{DFT}} \leq 50$ meV~atom$^{-1}$ yields 438 distinct structures across 30 compositions. To rapidly evaluate ionic conductivity, a single MD simulation is performed for each selected configuration using SevenNet-0 at 1000 K, and only those with conductivities above 101 mS~cm$^{-1}$ are retained, following Ref.~\cite{Ceder_CS_2022}. For these structures, we conduct comprehensive MD simulations with 3–5 independent runs at each temperature (800, 900, 1000, 1100, and 1200 K). Ionic conductivities at 300 K are extrapolated using the Arrhenius relationship, as described in Ref.~\cite{Mo_Arrhenius_2018}. This screening identifies 92 candidates with extrapolated room-temperature ionic conductivities exceeding 0.1 mS~cm$^{-1}$. Finally, for further validation, DFT-based MD simulations are carried out on 19 structures within 10 meV~atom$^{-1}$ of the lowest energy for each composition ($\Delta E^{{\rm{DFT}}}$), ultimately yielding 15 candidate structures with room-temperature conductivities greater than 0.1 mS~cm$^{-1}$. For the final candidates, we confirm that the conductivity values are not significantly affected by cell-size dependence (Fig.~S11).

\begin{figure*}[h]
\centering
  \includegraphics[width=13cm]{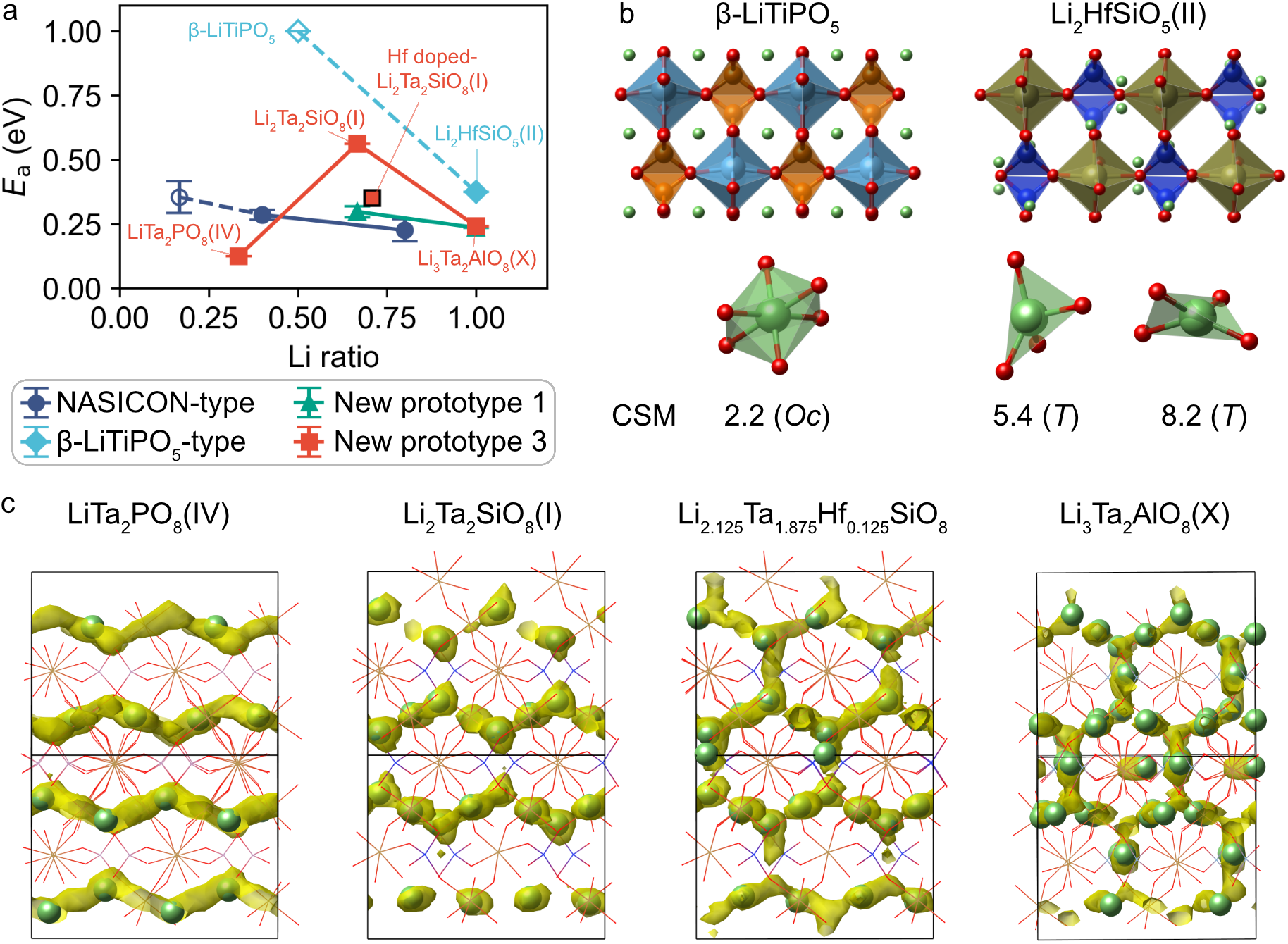}
  \caption{Effect of the Li ratio on ionic conductivity. (a) Activation energy as a function of Li ratio for NASICON-type, $\beta$-\ch{LiTiPO5}-type, new prototype 1, and new prototype 3. Markers indicate the average values of activation energy with the same framework and Li ratio, and error bars indicate the standard deviations. Activation energy values calculated with DFT are shown as filled markers, whereas experimentally reported values are shown as empty markers (empty diamond: $\mathrm{\beta}$-\ch{LiTiPO5}-type\cite{Robertson_LiTiPO5_Ea_1994}; empty circle: NASICON-type\cite{Aono_LTP_Ea4_1991, Takada_LTP_Ea2_2001, Arbi_LTP_Ea3_2006, Luo_LTP_Ea1_2023}). The corresponding compounds for the NASICON-type and new prototype 3 structures are indicated in the figure, and details for the other types are provided in Table~S4. (b) Structure and local environments of Li ions in $\mathrm{\beta}$-\ch{LiTiPO5} and \ch{Li2TiSiO5}. CSM values and the site type (\textit{Oc} for octahedron and \textit{T} for tetrahedron) for each Li environment are provided below. (c) Li conduction pathways at 800 K for \ch{LiTa2PO8}, \ch{Li2Ta2SiO8}, Hf-doped \ch{Li2Ta2SiO8}, and \ch{Li3Ta2AlO8}.}
  \label{Fig6}
\end{figure*}

Fig.~\ref{Fig5} presents the 15 identified candidates, with detailed properties provided in Table~\ref{Table3}. Several of these candidates share frameworks with known compounds, including \ch{Li2HfSiO5}(I) (LiTaSiO$_5$-type), \ch{Li2HfSiO5}(II) ($\beta$-LiTiPO$_5$-type), \ch{LiNb2PO8}(III) (LiTa$_2$PO$_8$-type), and three NASICON-type structures: \ch{Li4Hf2Si3O12}(I), \ch{Li4Zr2Si3O12}(I), and \ch{Li2Nb2Si3O12}(II). Note that the Roman numerals in parentheses indicate the stability order among configurations of the same composition, corresponding to their energy ranking relative to the ground state (with I denoting the ground state). In addition, TOPIC discovers materials with novel frameworks. For example, TOPIC reveals a distinct structural feature, commonly observed in \ch{Li2HfSi2O7}, \ch{Li2ZrSi2O7}, and \ch{Li3ScSi2O7}, consisting of \ch{Si2O7} polyanions interconnected by \textit{Oc}O$_6$ octahedra. The conduction pathways in these materials are found to be quasi two-dimensional (see Fig.~S12.a). Another structural type, featuring similar \ch{Si2O7} and \textit{Oc}O$_6$ connectivity but a slightly different atomic arrangement, is shared by \ch{Li2HfSi2O7} and \ch{Li2ZrSi2O7} (see blue dashed box in Fig.~\ref{Fig5}b), leading to the three-dimensional conduction pathways (see Fig.~S12.b). \ch{LiNb2PO8}(II) adopts a framework identical to that of metastable \ch{LiTa2PO8}, newly identified by TOPIC in the previous subsection (Fig.~\ref{Fig2}b). Li conduction pathways of three additional structures with previously unreported frameworks—\ch{Li2Nb2Si3O12}(I), \ch{LiNbSi2O7}(VIII), and \ch{Li3Nb3Si2O13}(X)—are also shown in Fig.~S12. We find that several candidates in this set exhibit high ionic conductivities above 1 mS~cm$^{-1}$. In particular, the NASICON-type \ch{Li4Hf2Si3O12} shows an ionic conductivity of 14.09 mS~cm$^{-1}$, which is comparable to that of liquid electrolytes and to the highest values reported for SSEs, such as \ch{LiNbOCl4} and \ch{LiTaOCl4}.\cite{Tanaka_oxyhalide_SSE_2023}

In addition to the final candidates, we find that several structural frameworks (e.g., NASICON and new prototypes 1 and 3) recur across new materials generated by TOPIC that satisfy $E_{\rm hull}^{\rm DFT} \leq 50$ meV, albeit with varying Li ratios (see Fig.~\ref{Fig6}a). Comparing the ionic conductivities of compounds sharing identical frameworks but different Li content can provide crucial insights for discovering novel SSEs. In particular, stuffing Li into SSEs is a well-recognized strategy for enhancing Li-ion conductivity, as it lowers the migration barrier through increased Li–Li repulsion and the resulting structural distortions.\cite{Ceder_CS_2022, Mo_LiAlSiO4_LiTaSiO5_2017} To examine whether this effect is also present in our candidates, we compare the activation barriers as a function of Li content for materials sharing the same framework. We analyze the NASICON-type, $\beta$-LiTiPO$_5$-type, and new prototypes 1 and 3, as shown in Fig.~\ref{Fig6}a. In all frameworks except new prototype 3, the activation barriers decrease as the Li ratio increases, which is consistent with previous theory. For instance, the activation barrier of synthesized $\beta$-\ch{LiTiPO5} exceeds 1 eV, as measured experimentally,\cite{Robertson_LiTiPO5_Ea_1994} whereas its analogue \ch{Li2HfSiO5}(II), which has a higher Li ratio owing to the smaller valence of Si compared to P, exhibits a much lower barrier of 0.38 eV. Continuous Symmetry Measure (CSM)\cite{Stefano_CSM_2019} analysis supports this trend: Li sites in $\beta$-\ch{LiTiPO5} show modest distortion (CSM = 2.2), while those in \ch{Li2HfSiO5}(II) show significantly larger distortions (CSM = 5.4 and 8.2). In the atomistic structures in Fig.~\ref{Fig6}b, we also see that the Li sites are symmetric in $\beta$-\ch{LiTiPO5}, whereas they are distorted in \ch{Li2HfSiO5}(II). The same CSM trends depending on the Li ratio are also observed in the NASICON-type and new prototype 1 structures (see Table~S5).

However, structures belonging to prototype 3 do not follow this trend. \ch{LiTa2PO8}, despite having the lowest Li ratio, exhibits the lowest activation barrier because it contains quasi-one-dimensional channels that provide efficient conduction pathways. By contrast, in \ch{Li2Ta2SiO8} (Li ratio = 0.667), the one-dimensional channels become overcrowded with Li ions, blocking not only the stable sites but also the transition-state sites along the migration pathway, which suppresses conductivity. Increasing the Li ratio further, as in \ch{Li3Ta2AlO8} (Li ratio = 1.0), restores conductivity by opening alternative three-dimensional pathways. Similarly, introducing a small amount of aliovalent doping in \ch{Li2Ta2SiO8}—for example, substituting Ta with Hf—creates new three-dimensional pathways and thereby lowers the activation barriers from 0.562 eV to 0.352 eV.


\begin{table*}
\small
  \caption{\ Properties of candidate structures found by TOPIC. Energy above hull (${E_{\mathrm{hull}}^{\mathrm{DFT}}}$), energy difference from the lowest-energy structure at the given composition($\Delta{E^{\mathrm{DFT}}}$), bandgap (${E_{\mathrm{g}}}$), activation energy ($E_\mathrm{a}$), and ionic conductivity at 300 K (${\sigma_{\mathrm{300K}}}$) are provided. For ${\sigma_{\mathrm{300K}}}$, the mean value ± standard deviation is shown in parentheses.}
  \label{Table3}
  \begin{tabular*}{\textwidth}{@{\extracolsep{\fill}}llllll}
    \hline
    Formula & ${E_{\mathrm{hull}}^{\mathrm{DFT}}}$ & $\Delta E^{\mathrm{DFT}}$ & ${E_{\mathrm{g}}}$ & ${E_{\mathrm{a}}}$ & ${\sigma_{\mathrm{300K}}}$\\
    & (meV~atom$^{-1}$) & (meV~atom$^{-1}$) & (eV) & (eV) & (mS~cm$^{-1}$) \\
    \hline
    \ch{Li2HfSi2O7} (I) & 12 & 0 & 6.8 & 0.319 ± 0.070 & 0.392 (0.026, 5.879)\\
    \ch{Li2HfSi2O7} (III) & 18 & 6.2 & 6.9 & 0.234 ± 0.082 & 2.388 (0.100, 56.801)\\
    \ch{Li2HfSiO5} (I) & 19 & 0 & 7.0 & 0.299 ± 0.020 & 1.127 (0.516, 2.463) \\
    \ch{Li2HfSiO5} (II) & 19 & 0.2 & 6.9 &  0.375 ± 0.063 & 0.113 (0.010, 1.276)\\
    \ch{Li4Hf2Si3O12} (I)  & 21 & 0 & 6.5 & 0.183 ± 0.016 & 14.090 (7.569, 26.229)\\
    \ch{LiNb2PO8} (II) & 25 & 9.3 & 4.2 & 0.353 ± 0.080 & 0.092 (0.004, 1.990)\\
    \ch{LiNb2PO8} (III) & 26 & 10.0 & 3.8 & 0.201 ± 0.034 & 7.820 (2.079, 29.416)\\
    \ch{Li2ZrSi2O7} (VI) & 28 & 2.1 & 6.3 & 0.277 ± 0.017 & 1.245 (0.657, 2.356)\\
    \ch{Li2ZrSi2O7} (VII) & 28 & 2.4 & 6.5 & 0.288 ± 0.067 & 0.553 (0.042, 7.335)\\
    \ch{Li3ScSi2O7} (II) & 28 & 6.0 & 6.7 & 0.236 ± 0.075 & 3.375 (0.188, 60.750)\\
    \ch{LiNbSi2O7} (VIII) & 29 & 9.2 & 5.0 & 0.328 ± 0.019 & 0.161 (0.078, 0.332)\\
    \ch{Li2Nb2Si3O12} (I)  & 31 & 0 & 4.5 & 0.301 ± 0.085 & 0.232 (0.009, 6.211)\\
    \ch{Li2Nb2Si3O12} (II)  & 32 & 0.6 & 4.8 & 0.267 ± 0.023 & 3.006 (1.245, 7.256)\\
    \ch{Li4Zr2Si3O12} (I) & 34 & 0 & 6.0 & 0.270 ± 0.006 & 0.756 (0.599, 0.955)\\
    \ch{Li3Nb3Si2O13} (X) & 44 & 9.7 & 3.9 & 0.274 ± 0.021 & 2.182 (0.959, 4.961)\\
    
    \hline
  \end{tabular*}
\end{table*}

\section{Discussion}\label{sec3}

In this study, we have developed TOPIC for the prediction of CS-topology crystal structures in Li-oxide SSEs. For a single quaternary composition containing approximately 100 atoms per unit cell, the prediction of the lowest-energy structure requires ~4 days of computational time. Given the intrinsic complexity of quaternary oxides and their large unit cells, this performance illustrates the efficiency of TOPIC, which requires only about twice the computational cost of earlier methods applied to simpler ternary oxides (about 50 atoms per unit cell).\cite{Hwang_JACS_2023}

By employing TOPIC, we have identified promising Li-oxide SSEs with high Li conductivity, as summarized in Table~\ref{Table3}. We have discovered a diverse set of candidate materials exhibiting previously unreported structural prototypes and elemental combinations. Interestingly, while high-performance solid electrolytes containing silicon have been rarely reported in previous studies, our screening reveals that 13 out of 15 low-energy Si-containing compositions exhibit high predicted ionic conductivity. Among these, five materials contain Hf, which is also uncommon among reported. Notably, \ch{Li4Hf2Si3O12} exhibits a high Li-ion conductivity of 14.09 mS~cm$^{-1}$, along with a wide band gap of 6.5 eV predicted using the accurate HSE06 method. In particular, this compound is predicted to be the lowest-energy structure among those with the same composition, implying relatively high synthesizability. Furthermore, we report various novel structural prototypes in this study (Fig.~\ref{Fig5}), which can serve as a materials library for designing new compounds through modifications such as phase distortions\cite{Awaka_Kijima_Hayakawa_Akimoto_2009} and aliovalent doping.\cite{Hitz_LLZO_Y_doped_2013, Adams_LLZO_Ta_doped_2011} For instance, we test the effectiveness of an aliovalent doping strategy on \ch{Li2Ta2SiO8}, whose lowest-energy structure with a novel framework exhibits low ionic conductivity (0.0004 mS~cm$^{-1}$) at 300 K. With Hf doping (Li$_{2.125}$Ta$_{1.875}$Hf$_{0.125}$SiO$_8$), the activation energy decreases by 0.210 eV, and the ionic conductivity at 300 K increases by more than an order of magnitude (0.193 mS~cm$^{-1}$). We also observe that doping induces a transition from a quasi-1D path in the undoped structure to a 3D Li conduction network, highlighting that doping creates additional conduction pathways (Fig.~\ref{Fig6}c). This is consistent with the transition to a 3D conduction pathway that occurs with increasing Li content across different element sets, as described in the previous section.

Even if this work conducts a comprehensive search of quaternary compositions composed of octahedral and tetrahedral polyhedra, this does not encompass the full chemical space of oxide SSEs. For example, the present screening does not include elements forming 3-coordinated polyhedra (e.g., SeO$_3$ in \ch{LiGa(SeO3)2}\cite{Ceder_CS_2022}) or garnet-type frameworks with La atoms such as LLZO. Since we have demonstrated that these frameworks can also be correctly predicted by TOPIC with a slight modification, future studies will expand to searching these chemical spaces as well. Furthermore, quinary compositions and doped materials may also exhibit high ionic conductivity. The structure libraries constructed in this work provide a valuable foundation for pursuing such extensions. 
The discovery of these novel compositions and prototypes is enabled by the ability of the TOPIC algorithm to directly explore the potential energy surface, thereby overcoming the limitations of conventional template-based methods.\cite{Walsh_datamining_limitation_2017}

\section{Conclusions}\label{sec4}
In summary, we develop a new structure prediction algorithm, TOPIC, designed to efficiently generate oxide SSE materials with CS frameworks. We validate that TOPIC successfully reproduces known oxide SSEs (up to ~200 atoms per unit cell) as well as previously unreported polymorphs. By integrating a pretrained MLIP, SevenNet-0, we explore uncharted chemical spaces and identify several promising candidates. Our analysis reveals that the Li ratio plays a critical role in determining the stability of CS frameworks and provides a useful descriptor for screening. We analyze the effect of Li content on ionic conductivity from the discovered structures, providing guidance for establishing design principles. Overall, TOPIC offers a scalable route to explore uncharted chemical space for next-generation oxide electrolytes.

\section{Methods}\label{sec5}
\subsection{TOPIC algorithm}
\textbf{Random structure generation for cation sublattices.}
Cation sublattices are generated under specified space groups using the RandSpg package,\cite{Avery_randSpg_2017} which randomly assigns lattice vectors and Wyckoff positions while preserving the designated symmetry. The symmetry group of each cell is chosen randomly. The mean cell volumes of the generated structures are referenced to the density of the amorphous phase obtained from DFT melt–quench simulations (see below). Volumes are selected within $90$–$120$\% of the mean cell volume, with each lattice parameter constrained to $40$–$250$\% of the cubic root of the mean cell volume. Lattice angles are restricted to $60^{\circ}$–$120^{\circ}$. Distance constraints are applied to all cation-–cation pairs, by imposing a minimum separation equal to the values observed for the corresponding pairs during the melt–quench–annealing trajectories. This procedure ensures chemically reasonable configurations and avoids unphysical short interatomic distances.

\textbf{Oxygen placement.} After generating random cation sites, oxygen atoms are placed at the midpoints of cation pairs. Two different algorithms are employed to generate cation pairings (virtual bonds) that satisfy both the CS bond topology and coordination number constraints. (1) In the first algorithm, the bond topology is generated by sequentially forming virtual bonds for each cation according to its predefined coordination number. Specifically, the order of cations is first randomly determined. For each cation in this order, connections are made to the nearest neighboring cations until its coordination number is satisfied. If a virtual bond has already been assigned by another cation, the bond is retained, and additional connections are made preferentially to the nearest unsaturated neighbors. If the nearest cation has already reached its coordination number limit, the next nearest cation is chosen, and this process is repeated until all cations satisfy their coordination numbers. When this process fails (e.g., by forming isolated loops), a new random cation order is selected, and the procedure is retried up to 1000 times. If all attempts fail, or if the generated structure succeeds but has P1 symmetry, a new random cation configuration is generated and the algorithm is repeated until success. Once the bond topology is completed, oxygen atoms are placed at the midpoints of the virtual bonds. (2) In the second algorithm, all cation pairs with short separations that can form virtual bonds are first enumerated, and only those structures in which a CS bond topology is generated are selected. Specifically, all cation pairs with separations shorter than twice the cation–O distance (the first radial distribution function (RDF) valley from the MD simulations) are considered, and oxygen atoms are placed at the midpoints accordingly. After assigning oxygen atoms, the coordination number of each cation is checked. If any cation does not satisfy its predefined coordination number, the cation structures are regenerated under the same space group, and this process is repeated up to 100 trials. If all attempts fail, a new space group number is randomly selected and the process is repeated until success. Compared with the first method, this approach generally preserves the symmetry of the initial cation arrangement, and the resulting structures therefore tend to exhibit higher symmetry. In practice, structures are generated such that 50\% are obtained from the first method and 50\% from the second.

\textbf{Bond-length optimization with classical potentials.} Oxygen atoms are initially placed at the midpoints of the virtual bonds to satisfy the CS bond topology. However, the resulting cation--O bond lengths are not optimal. Using LAMMPS package,\cite{Thompson_LAMMPS_2022} the structures are optimized using a Lennard--Jones (LJ) potential,
\[
V(r) = 4\varepsilon \left[ \left( \tfrac{\sigma}{r} \right)^{12} - \left( \tfrac{\sigma}{r} \right)^{6} \right],
\]
which is selectively applied to the virtual bonds between cations and oxygen atoms. Note that the LJ potentials are not applied to all pairs of atoms, but are selectively applied to O–-cation pairs connected by virtual bonds. Here, $\sigma$ is set to the position of the first peak of the cation--O RDF obtained from amorphous structures generated by DFT-MD simulations for the training set, divided by $2^{\tfrac{1}{6}}$, ensuring that the equilibrium bond length coincides with the RDF peak. The $\varepsilon$ value is set to 3.0 eV for all cases. To prevent the formation of highly unphysical structures, additional harmonic repulsion potentials are introduced,
\[
E = K (R - R_c)^2,
\]
whenever two atoms are closer than the cutoff distance $R_c$ (fix restrain command in LAMMPS). The $K$ value is set to 0.2 eV~\AA{}$^{-2}$ for all cases. These repulsion terms are applied between O--O pairs, between cation--cation pairs, and also between cations and O atoms belonging to different polyhedra. For the cutoff distances of O--O and cation--cation pairs, the shortest distances observed in melt--quench--annealing MD trajectories are used. In the case of cation--oxygen repulsion across different polyhedra, the cutoff is set to the position of the first valley in the cation--O RDF of the amorphous structure. 

\textbf{Optimizing frameworks.} After optimization with the LJ potential, the code checks whether the structure satisfies the desired coordination numbers and CS bond topology. If this condition is not met, the structure is discarded; otherwise, it is further relaxed with the MLIPs trained without labeling the Li atoms. The details of training set generation and model architectures are described below. The structure is first optimized using an MLIP with a cutoff of 3 \AA{}, and then further relaxed using a model with a cutoff of 6.0 \AA{}. At each stage, structures that do not preserve the CS bond topology or bond-length condition after relaxation are discarded.

\textbf{Li insertion process.} We use Voronoi tessellation based on oxygen atom positions\cite{Ceder_CS_2022} to enumerate possible Li sites using SciPy package.\cite{Virtanen_SciPy_2020} Among the Voronoi nodes, those located within 1 \AA{} of non-Li cations are discarded. The remaining nodes are clustered if they are within 1 \AA{} of each other, and represented by the average position. To obtain stable configurations, Li atoms are first placed in the largest free spaces and the structures are relaxed using MLIPs (trained for all elements including Li). If the structure collapses such that the coordination numbers of the non-Li cations are altered, Li atoms are inserted at random candidate sites and the structures are checked again. If no structure satisfying the conditions is obtained within 10 trials, the framework is considered unstable upon Li insertion and discarded. If the framework remains stable, 0 K Monte Carlo (MC) simulations are performed. In each MC step, one Li atom is randomly selected and its position is changed to one of the candidate Li sites. After relaxation with MLIPs, the bond topology and coordination numbers are checked. If the conditions are not satisfied or the energy increases, the move is rejected. After 100 MC steps, the lowest-energy structure is selected.

\textbf{Final energy evaluations.} After generating candidate structures with the above processes, the final candidates within an energy window of 100 meV~atom$^{-1}$ from the MLIP calculations are selected and then recalculated using DFT. To improve efficiency, we adopt a two-step procedure. First, one-shot calculations are performed for all candidate structures using a $k$-point grid with the same $k$-spacing as in the DFT melt–quench simulations. Next, structures within an energy window of 50 meV~atom$^{-1}$ from the one-shot DFT results are selected and relaxed with the AMP$^2$ package\cite{Youn_AMP2_2020} to obtain accurate energies. Convergence criteria for the $k$-point tests are set to 3 meV~atom$^{-1}$ for energy and 10 kbar for pressure. The cutoff energy for all DFT calculations is set to 520 eV.

\textbf{Modification of TOPIC algorithm for \ch{Li7La3Zr2O12}.}
Generating the \ch{(ZrO6/2)2(LiO4/2)3} framework follows the standard TOPIC workflow: sampling random cation sublattices and optimizing bond lengths with LJ potentials. During framework optimization, however, we use only the 3 Å–cutoff MLIP, because the 6 Å–cutoff model exhibits roughly twice the force error of our validation systems, whereas the 3 Å model is only ~40\% higher. During the La-insertion MC, we evaluate single-point energies rather than performing relaxations, in order to avoid artifacts arising from the non-stoichiometric conditions. Finally, we remove the Li atoms within the framework and reinsert Li atoms via MC simulations, yielding the final \ch{Li7La3Zr2O12} structures.

\subsection{Machine learning interatomic potential training}
\textbf{MLIP details.} We employ Behler–Parrinello neural network potentials,\cite{Behler_Parrinello_2007} with symmetry function vectors as descriptors, using the SIMPLE-NN package.\cite{Lee_SIMPLE-NN_2019} For each atomic pair, 8 radial and 18 angular symmetry functions are constructed with identical cutoff radii applied to both. Model training is continued until the root-mean-square errors (RMSEs) of the validation set fails to decrease by at least 0.5 meV~atom$^{-1}$ (energy), 0.01 eV~\AA$^{-1}$ (force), or 0.5 kbar (stress) over 50 epochs. The average validation RMSEs are: (1) 39 meV~atom$^{-1}$ (energy), 2.0 eV~\AA$^{-1}$ (force), and 59 kbar (stress) for the 3 \AA cutoff models excluding Li atoms; (2) 14 meV~atom$^{-1}$, 0.87 eV~\AA$^{-1}$, and 44 kbar for the 6 \AA cutoff models excluding Li atoms; and (3) 4 meV~atom$^{-1}$, 0.33 eV~\AA$^{-1}$, and 8 kbar for the 6 \AA cutoff models including all atoms. The neural networks consists of two hidden layers with 30 nodes each. Input vectors are decorrelated using principal component analysis (PCA) and subsequently whitened to accelerate convergence.\cite{Yoo_Atomic_energy_mapping_2019} Model weights are optimized using the Adam algorithm,\cite{Kingma_Adam_2014} with a batch size of 20. To mitigate overfitting, L2 regularization is incorporated into the loss function, and 10\% of the dataset is reserved for validation.

\textbf{Generating training set.} Disordered structures for the training set are generated using melt–-quench--annealing MD simulations with DFT calculations. Initial configurations are constructed by randomly placing atoms within a volume estimated from their atomic radii. The system is first equilibrated at 4500 K for 3 ps, followed by melting at an empirically determined melting temperature for 8 ps. The melting temperature is estimated as the point where the diffusion coefficient of all atomic species exceeds $4 \times 10^{-9}\,\mathrm{m^2~s^{-1}}$, a common characteristic value for liquid metals and ceramics.\cite{Protopapas_Liquid_D_threshold_1973, Nowok_D_threshold_1996} From this molten state, the system is quenched at a cooling rate of 200 K ps$^{-1}$ and subsequently annealed at 500 K for 4 ps, yielding amorphous configurations suitable for MLIP training and providing preliminary structural information.

\textbf{Density functional theory calculations.} All DFT calculations are performed using the Vienna Ab initio Simulation Package (VASP)\cite{Kresse_VASP_1996} with projector augmented-wave (PAW) pseudopotentials\cite{Blöchl_PAW_1994} and the Perdew–Burke–Ernzerhof (PBE) functional\cite{Perdew_GGA_1996} for the exchange–correlation energy. The cutoff energies are determined from convergence tests on premelted structures, ensuring that the total energies, forces, and stress tensors converge within 20 meV~atom$^{-1}$, 0.3 eV~\AA$^{-1}$, and 10 kbar, respectively. MLIPs trained on melt–quench–anneal trajectories that satisfied these thresholds are shown to be sufficiently accurate for crystal-structure prediction, successfully generating low-energy structures in the process.\cite{Kang_SPINNER_2022, Hwang_JACS_2023} A single $k$-point, either $\Gamma$ or ($\frac{1}{4}$, $\frac{1}{4}$, $\frac{1}{4}$),\cite{Baldereschi_1972} is used for the Brillouin-zone integration, with the choice determined by the same convergence tests as for the cutoff energies. Band gap values of the final candidates are evaluated by one-shot hybrid functional calculations (HSE06 functional)\cite{Heyd_HSE06_2003} on the PBE-optimized structures using the AMP$^2$ package.\cite{Kim_AMP2_bandgap_2020}

\subsection{SPINNER} 
We use SPINNER to generate structures without bond topology constraints.\cite{Kang_SPINNER_2022} Structures are produced through random seeding, permutation, and lattice mutation in ratios of 70, 20, and 10\%, respectively. The population size of each generation is capped at 300, and the search proceeds for 200 generations. In each generation, structures within a 100 meV atom$^{-1}$ window from the lowest-energy structure of the previous generation are selected for mutation operations. After the entire run, the final candidate structures within the lowest 50 meV atom$^{-1}$ energy window are retained. During structural relaxations, we use the same MLIPs as in TOPIC, trained with a 6 \AA{} cutoff radius and including all atoms.

\subsection{Li ionic conductivity calculation}
We follow the method described in ref.\cite{Mo_Arrhenius_2018}. MD simulations up to 50 ps (both DFT and MLIP) are conducted 3-5 times, and a time-averaging method is used to calculate the mean squared displacement (MSD), defined as follows:

\begin{equation}
\mathrm{MSD}(\tau) = \frac{1}{N_{\mathrm{Li}}} \sum_{i=1}^{N_{\mathrm{Li}}} \left\langle \left| \mathbf{R}_i(t+\tau)-\mathbf{R}_i(t) \right|^2 \right\rangle _t
\end{equation}
where $N_{\mathrm{Li}}$ is the number of Li atoms, $\mathbf{R}_i(t)$ is position vector of Li atom $i$ at time $t$, and $\tau$ is the lag time. The notation $\langle \cdots \rangle_t$ denotes an average over all permissible time origins $t$ within a trajectory. The diffusion coefficient at each temperature is calculated from the Einstein relation:

\begin{equation}
D_{\mathrm{Li}} = \lim_{t\to\infty}\frac{\mathrm{MSD}(t)}{6t} = D_0 \exp \left( -\frac{E_{\mathrm a}}{k_{\mathrm B} T} \right)
\end{equation}
where $D_0$ is the pre-factor of the diffusion coefficient, $E_a$ is the activation energy of diffusion, ${k_{\mathrm B}}$ is Boltzmann constant, and $T$ is temperature. To obtain a linear region of MSD, we remove the initial 2 ps of MSD and beyond 50\% of the total simulation time. The activation energy and ionic conductivity are obtained by fitting the diffusion coefficients to the Arrhenius equation in the temperature range of 800–1200 K:

\begin{equation}
\log (D_{\mathrm{Li}}) = - \frac{E_{\mathrm a} }{k_{\mathrm B} T} + C
\end{equation}
where $C$ is a temperature-independent constant. We apply weights of the inverse square of the standard deviation of log($D_{\mathrm{Li}}$) to each data point during the fitting to handle the statistical uncertainty.

The ionic conductivity is then calculated from the Nernst–Einstein relation:

\begin{equation}
\label{eqn:Nernst-Einstein}
\sigma_{\mathrm{Li}} = \frac{N_{\mathrm{Li}} z^2 e^2}{V k_{\rm B} T}D_{\mathrm{Li}}
\end{equation}
where $e$ is the elementary charge and $V$ is the volume of the simulated cell. We set $z=1$ for Li ions.

For visualization of Li conduction pathways, the pymatgen-analysis-diffusion package is employed,\cite{Deng_Pymatgen_diffusion_2017, Zhu_Pymatgen_diffusion_anaylsis_2015} and plotted isosurfaces of Li probability density at a threshold of $0.001~P_{\mathrm{max}}$, where $P_{\mathrm{max}}$ denotes the maximum value of the Li-ion probability density distribution.

While SevenNet-0 shows good predictive performance for materials with known structures,\cite{Loew_phonon_7net_2025, Riebesell_phase_stability_7net_2025, Csányi_thermal_conductivity_7net_2025} its reliability for newly discovered prototypes remains uncertain. To assess its applicability to evaluate Li-ion transport, we compare diffusion coefficients predicted by SevenNet-0 with those obtained from DFT (Fig.~S13). For compounds with known prototypes, the model achieves a mean percentage error (MPE) of 8\% and a mean absolute percentage error (MAPE) of 23\%. In contrast, for the novel frameworks discovered in this study, SevenNet-0 tends to overestimate the diffusion coefficients, with an MPE of 32\% and a MAPE of 42\%, likely arising from a softening of the potential energy surface relative to DFT.\cite{Deng_force_softening_2025, Kim_FT_SevenNet_2025} Nevertheless, these errors do not alter the order of magnitude of the predicted conductivities. Thus, SevenNet-0 remains suitable for high-throughput screening of oxide-based Li-ion solid electrolytes.

\subsection{Aliovalent doping}
Voronoi nodes are considered as candidate Li sites, as described above. For each system, 50 structures with random cation substitutions and Li stuffing are generated. Their energies are evaluated using SevenNet-0, and the lowest-energy structure is further relaxed with SevenNet-0. Although the selected structures may not correspond to the absolute lowest-energy configurations, they are expected to represent the doped systems reasonably well, as the ionic conduction characteristics are reproduced in line with previous reports (e.g., Ga-doped \ch{LiTiPO5} and P-doped \ch{Li2Mg2(SO4)3}, see Table~S2).

\section{Data availability}
The energy values and corresponding structure files for the discovered structures are provided separately as Supplementary Information. 

\section{Code availability}
The main part of TOPIC is opened at (https://github.com/kang1717/TOPIC). SPINNER for carrying evolutionary structure searches is available at (https://github.com/MDIL-SNU/SPINNER).\cite{Kang_SPINNER_2022} SIMPLE-NN for training MLIPs is available at (https://github.com/MDIL-SNU/SIMPLE-NN\_v2).\cite{Lee_SIMPLE-NN_2019} The SevenNet code utilized in this study is available in the project GitHub repository (https://www.github.com/MDIL-SNU/SevenNet).\cite{Park_SevenNet_2024}

\section{Acknowledgements}\label{sec5}

This research was supported by the Nano \& Material Technology Development Programs through the National Research Foundation of Korea (NRF) funded by Ministry of Science and ICT (RS-2024-00407995 and RS-2024-00450102). The computations were carried out at the Korea Institute of Science and Technology Information (KISTI) National Supercomputing Center (KSC-2025-CRE-0110).

\section{Author contributions}
S.K. conceived the original idea of topology constraints and developed the prototype code. S.Hw. developed the main algorithm of TOPIC and carried out the major tasks, including code implementation, MLIP training, performance tests, screening, and data analysis. J.L. trained of MLIPs for a subset of compositions. S.K., Y.K., and S.Ha. supervised the project and provided discussions. All authors contributed to drafting the manuscript.

\bibliography{sn-bibliography}
\end{document}